\documentclass[11pt]{article}

\usepackage{palatino}
\usepackage{mathpazo}
\usepackage{stmaryrd}

\usepackage{gastex}
\usepackage{amsfonts}
\usepackage{amssymb}
\usepackage{amsmath}
\usepackage{latexsym}
\usepackage{amsthm}
\usepackage{eepic}
\usepackage{sectsty}
\usepackage[usenames]{color}
\usepackage{hyperref}

\hypersetup{pdfpagemode=UseNone}

\theoremstyle{definition}

\newcommand{\tinyspace}{\mspace{1mu}}

\newcommand{\abs}[1]{\left\lvert\tinyspace #1 \tinyspace\right\rvert}

\newcommand{\norm}[1]{\left\lVert\tinyspace #1 \tinyspace\right\rVert}

\newcommand{\ket}[1]{\left| #1 \right\rangle}
\newcommand{\bra}[1]{\left\langle #1 \right|}

\renewcommand{\outer}[1]{|#1\rangle\langle#1|}

\newcommand{\tr}{\operatorname{Tr}}

\def\I{\mathbb{1}}

\newcommand{\setft}[1]{\mathrm{#1}}
\newcommand{\lin}[1]{\setft{L}\left(#1\right)}
\newcommand{\density}[1]{\setft{D}\left(#1\right)}

\def\real{\mathbb{R}}
\def\natural{\mathbb{N}}

\newenvironment{mylist}[1]{\begin{list}{}{
	\setlength{\leftmargin}{#1}
	\setlength{\rightmargin}{0mm}
	\setlength{\labelsep}{2mm}
	\setlength{\labelwidth}{8mm}
	\setlength{\itemsep}{0mm}}}
	{\end{list}}

\def\X{\mathcal{X}}
\def\Y{\mathcal{Y}}

\def\W{\mathcal{W}}

\def\({\left(}
\def\){\right)}

\begin{document}

\title{\bf\LARGE
  Adaptive versus non-adaptive strategies for\\
  quantum channel discrimination}

\author{
  Aram W.\ Harrow$^\ast$
  \hspace{7mm}
  Avinatan Hassidim$^\dagger$
  \hspace{7mm}
  Debbie W.\ Leung$^\ddagger$
  \hspace{7mm}
  John Watrous$^\ddagger$\\[3mm]
  $^\ast$%
  {\small\it Department of Mathematics, University of Bristol}\\[-1mm]
  {\small\it Bristol, United Kingdom}\\[1mm]
  $^\dagger$%
  {\small\it Center for Theoretical Physics, Massachusetts Institute
    of Technology}\\[-1mm]
  {\small\it Cambridge, Massachusetts, USA}\\[1mm]
  $^\ddagger$%
  {\small\it Institute for Quantum Computing, University of
    Waterloo}\\[-1mm]
  {\small\it Waterloo, Ontario, Canada}
}

\date{September 1, 2009}

\maketitle

\begin{abstract}
  We provide a simple example that illustrates the advantage of
  adaptive over non-adaptive strategies for quantum channel
  discrimination.
  In particular, we give a pair of entanglement-breaking channels that
  can be perfectly discriminated by means of an adaptive strategy that
  requires just two channel evaluations, but for which no non-adaptive
  strategy can give a perfect discrimination using any finite number
  of channel evaluations.
\end{abstract}
  
\section{Introduction} \label{sec:introduction}

This paper concerns the problem of {\it quantum channel discrimination}.
In this problem, two quantum channels $\Phi_0$ and $\Phi_1$ are fixed,
and access to one of the two channels is made available.
It is not known which of the two channels has been made available,
however, and the goal is to correctly identify which of $\Phi_0$ and
$\Phi_1$ it is.
Several papers, including
\cite{
  Acin01,
  AharonovKN98,
  ChildsPR00,
  ChiribellaDP08,
  D'ArianoPP01,
  DuanFY09,
  Hayashi08,
  Kitaev97,
  PianiW09,
  Sacchi05,
  Sacchi05b,
  WangY06,
  Watrous08},
have discovered many interesting aspects of quantum channel
discrimination.
There exist related topics in the study of quantum information theory,
including {\it quantum parameter estimation} (see, for instance
\cite{FujiwaraI03,ImaiH08,JiWDFY08} and the references therein), but
this paper will focus just on the specific problem of channel
discrimination.

A {\it discrimination strategy} for a quantum channel discrimination
problem is a step-by-step procedure consisting of channel evaluations,
along with quantum state preparations, operations, and measurements,
that attempts to output the identity of the given channel.
Generally speaking, one is typically interested in discrimination
strategies that satisfy certain natural constraints; with one
well-studied example being the discrimination strategies allowing a
{\it single evaluation} of the unknown channel.
An {\it optimal} discrimination strategy, among those satisfying a
given collection of constraints, is simply one that maximizes the
probability that the unknown channel is correctly identified, assuming
it is selected according to a fixed distribution that is known ahead
of time.

One interesting aspect of quantum channel discrimination is that the
use of an {\it ancillary} system is generally necessary for the
optimal discrimination of two quantum channels, assuming just a single
evaluation of the unknown channel is made available 
\cite{Kitaev97,AharonovKN98,D'ArianoPP01,KitaevSV02}.
In more precise terms, the optimal strategy to discriminate two
channels may require that one first prepares the input system to the
unknown channel in an entangled state with an ancillary system,
followed by a joint measurement of that channel's output together with
the ancillary system.
Even {\it entanglement-breaking} channels are sometimes better
discriminated through the use of an ancillary system, despite the fact
that their output systems must necessarily be unentangled with the
ancillary system after their evaluation \cite{Sacchi05b}.
There are two known special classes of channels that require no
ancillary system for optimal discrimination: the unitary channels
\cite{AharonovKN98,ChildsPR00} and the classical channels.

There is a striking possibility for quantum channel discrimination
problems that cannot occur in the classical setting.  
If a pair of classical channels cannot be perfectly distinguished with
one evaluation, then they cannot be perfectly distinguished with any
finite number of evaluations.
(This fact is easily proved, and a simple proof may be found later in
the paper.)
In contrast, it is possible for a pair of quantum channels to be
discriminated perfectly when multiple evaluations are available, but
not in the single evaluation case.
For example, this generally happens in the case of unitary channels
\cite{Acin01}.

Another interesting aspect of quantum channel discrimination is the
distinction between {\it adaptive} and {\it non-adaptive} strategies
when multiple uses of the unknown channel are made available.  
In an adaptive strategy, one may use the outputs of previous uses of
the channel when preparing the input to subsequent uses, whereas a
non-adaptive strategy requires that the inputs to all uses of the
given channel are chosen before any of them is evaluated.
It was found in \cite{ChiribellaDP08} that unitary channels are
insensitive to this distinction; adaptive strategies do not give
any advantage over non-adaptive strategies for unitary channel
discrimination.
In the same paper, a pair of {\it memory channels} was shown to
require an adaptive scheme for optimal discrimination, but the
question of whether or not there exist ordinary (non-memory) channels
with a similar property was stated as an open question.
Although an example of {\em three} channels that require
adaptive strategies for an optimal identification was presented in
\cite{WangY06}, we were not able to find any example of a pair of
(ordinary, non-memory) channels in the literature that require
adaptive strategies for optimal discrimination; and so the question
appears to have been unresolved prior to this work.

The purpose of the present paper is to demonstrate the necessity of
adaptive schemes for optimal quantum channel discrimination.
We do this by presenting an example of two quantum channels that can
be perfectly  discriminated given two adaptive channel evaluations,
but for which {\it no finite number} of non-adaptive channel
evaluations allows for a perfect discrimination.  
The channels in our example are {\it entanglement-breaking} channels,
which provides further evidence suggesting that entanglement-breaking
channels share similar properties to general quantum channels with
respect to channel discrimination tasks.
We note that a recent paper of Duan, Feng, and Ying \cite{DuanFY09}
has provided a criterion for the perfect discrimination of pairs of
quantum channels, as well as a general method to find adaptive
strategies that allow for perfect discrimination.
While no explicit examples were given in that paper, the existence of
pairs of channels with similar properties to those in our example is
implied.
Our example was, however, obtained independently from that paper, and
we hope that it offers some insight into the problem of quantum
channel discrimination that is complementary to \cite{DuanFY09}.

Finally, we note that a related (but weaker) phenomenon occurs in the
context of classical channel discrimination.
That is, there exist classical channels that can be better
discriminated by adaptive strategies than by non-adaptive strategies,
and we provide three simple examples illustrating this phenomenon.
While we suspect that similar examples illustrating the advantages of
adaptive discrimination strategies may be known to some researchers,
we did not find any in the literature.
That such examples exist is also interesting when contrasted with the
fact that adaptive strategies for classical channel discrimination
cannot improve the asymptotic rate at which the error probability
exponentially decays with the number of channel
uses~\cite{Hayashi08}.

\section{Preliminaries}

We will begin by summarizing some of the notation and terminology that
is used in the subsequent sections of the paper.
We will let $\X$, $\Y$ and $\W$ denote finite-dimensional complex
Hilbert spaces, which will typically correspond to the input, output,
and ancillary systems to be associated with channel discrimination
tasks.
The notation $\lin{\X,\Y}$ refers to the space of all linear operators
from $\X$ to $\Y$, $\lin{\X}$ is shorthand for $\lin{\X,\X}$, and
$\density{\X}$ refers to the set of all density operators on $\X$.
A similar notation is used for other spaces in place of $\X$ and $\Y$.
The identity operator on $\X$ is denoted~$\I_{\X}$.

For the example to be presented in the main part of the paper, we will
let $\X$ and $\Y$ be the spaces associated with two qubits and one
qubit, respectively.
The standard bases for these spaces are therefore
$\{\ket{00},\ket{01},\ket{10},\ket{11}\}$ and
$\{\ket{0},\ket{1}\}$.
As is common, we will also write
\[
\ket{+} = \frac{1}{\sqrt{2}}\ket{0} + \frac{1}{\sqrt{2}}\ket{1}
\quad\quad\text{and}\quad\quad
\ket{-} = \frac{1}{\sqrt{2}}\ket{0} - \frac{1}{\sqrt{2}}\ket{1},
\]
and we write tensor products of these states and standard basis
states in a self-explanatory way (e.g., $\ket{1+} = \ket{1}\ket{+}$).

A {\it quantum channel} is a linear mapping of the form
$\Phi : \lin{\X} \rightarrow \lin{\Y}$ that is both completely
positive and trace-preserving.
Every such quantum channel $\Phi$ can be expressed in Kraus form as
\[
\Phi(X) = \sum_{j = 1}^m A_j X A_j^{\ast}
\]
for some choice of linear operators $A_1,\ldots,A_m\in\lin{\X,\Y}$
satisfying the constraint
\[
\sum_{j = 1}^m A_j^{\ast} A_j = \I_{\X}.
\]
The identity channel mapping $\lin{\W}$ to itself is denoted
$\I_{\lin{\W}}$.

The distinguishability of two quantum channels
$\Phi_0,\Phi_1:\lin{\X}\rightarrow\lin{\Y}$ may be quantified by the
distance induced by the {\it diamond norm} (or 
{\it completely bounded trace norm})
\begin{equation}
\label{eq:diamond}
\norm{ \Phi_0 - \Phi_1}_{\diamond}
= \max_{\rho\in\density{\X\otimes\W}}
\norm{ (\Phi_0 \otimes \I_{\lin{\W}})(\rho) - (\Phi_1 \otimes
  \I_{\lin{\W}})(\rho)}_1
\end{equation}
where here $\W$ is assumed to have dimension at least that of $\X$.
This quantity represents the greatest possible degree of
distinguishability that can result by feeding an input state into the
two channels, allowing for the possibility that the input system is
entangled with an ancillary system.
Assuming that a bit $a\in\{0,1\}$ is uniformly chosen at random, the
quantity
\[
\frac{1}{2} + \frac{1}{4}\norm{\Phi_0 - \Phi_1}_{\diamond}
\]
represents the optimal probability to correctly determine the value
of $a$ by means of a physical process involving just a single
evaluation of the channel $\Phi_a$.
It therefore holds that $\Phi_0$ and $\Phi_1$ are perfectly
distinguishable using a single evaluation if and only if
$\norm{\Phi_0 - \Phi_1}_{\diamond} = 2$.

\section{Specification of the example and a perfect discrimination protocol}

We now describe our example of two quantum channels that are better
discriminated using an adaptive strategy than by any non-adaptive
strategy.
First, we will give an intuitive description of the channels.
The two channels, $\Phi_0$ and $\Phi_1$, both map two qubits to one
and operate as follows.

\begin{mylist}{2.5ex}
\item[$\bullet$]
Channel $\Phi_0$ measures the first input qubit with respect to the
standard basis.
If the result is 0, it outputs the state $\ket{0}$.
If the result is 1, it measures the second qubit with respect to the
standard basis.
If the result is 0, then it outputs 0, and if the result is 1, then it
outputs the completely mixed state $\I/2$.

\item[$\bullet$]
Channel $\Phi_1$ measures the first input qubit with respect to the
standard basis.
If the result is 0, it outputs the state $\ket{+}$.
If the result is 1, it measures the second qubit with respect to the
$\{\ket{+},\ket{-}\}$ basis.
If the result is $+$, then it outputs 1, and if the result is $-$, then it
outputs the completely mixed state $\I/2$.
\end{mylist}

The intuition behind these channels is as follows.
If the first input qubit is set to 0, then the output is a ``key''
state: $\ket{0}$ for channel $\Phi_0$ and $\ket{+}$ for the channel
$\Phi_1$.
If the first input is set to 1, and the second input qubit is the
channel's ``key'' state, then the channel identifies itself (i.e.,
$\Phi_0$ outputs 0 and $\Phi_1$ outputs~1).
If, however, the first input qubit is set to 1 and the second qubit's
state is orthogonal to the channel's ``key'' state, then the channel
outputs the completely mixed state.
This effectively means that the channel provides no information about
its identity in this case.

It is easy to discriminate these two channels with an adaptive strategy
that requires two uses of the unknown channel.
The following diagram describes such a strategy:
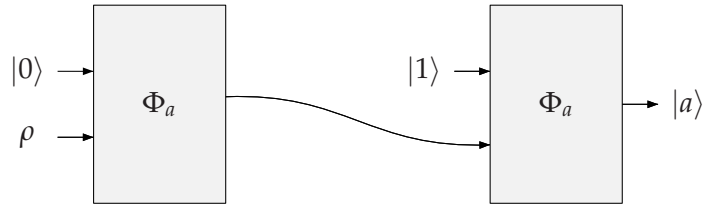
\begin{center}
  \unitlength=0.25pt
  \begin{picture}(1000, 400)(0,0)
    \gasset{Nmr=0}
    \node[Nw=200,Nh=300,fillgray=0.95](first)(200,200){$\Phi_a$}
    \node[Nw=200,Nh=300,fillgray=0.95](second)(800,200){$\Phi_a$}
    \node[Nframe=n,Nh=100](in1)(0,250){$\ket{0}$}
    \node[Nframe=n,Nh=100](in2)(0,150){$\rho$}
    \node[Nframe=n,Nh=100](in3)(600,250){$\ket{1}$}
    \node[Nframe=n,Nh=100](out)(1000,200){$\ket{a}$}
    \drawedge[eyo=50](in1,first){}
    \drawedge[eyo=-50](in2,first){}
    \drawedge[eyo=50](in3,second){}
    \drawbpedge[eyo=-50](first,10,320,second,190,320){}
    \drawedge(second,out){}
    \drawedge[eyo=-50](in2,first){}
    \drawedge[eyo=50](in3,second){}
    \drawbpedge[eyo=-50](first,10,320,second,190,320){}
    \drawedge(second,out){}
  \end{picture}
\end{center}
Here, the state $\rho$ input as the second qubit of the first channel
evaluation is arbitrary, as it is effectively discarded by both of the
channels when the first input qubit is set to $\ket{0}$.

In the interest of precision, and because it will be useful for the
analysis of the next section, we note the following formal
specifications of these channels.
It holds that
\[
\Phi_0(X) = \sum_{j = 1}^5 A_j X A_j^{\ast}
\quad\quad\text{and}\quad\quad
\Phi_1(X) = \sum_{j = 1}^5 B_j X B_j^{\ast}
\]
for
\begin{xalignat*}{5}
A_1 & = \ket{0}\bra{00}, &
A_2 & = \ket{0}\bra{01}, &
A_3 & = \ket{0}\bra{10}, &
A_4 & = \frac{1}{\sqrt{2}}\ket{0}\bra{11}, &
A_5 & = \frac{1}{\sqrt{2}}\ket{1}\bra{11}, \\
B_1 & = \ket{+}\bra{00}, &
B_2 & = \ket{+}\bra{01}, &
B_3 & = \ket{1}\bra{1+}, &
B_4 & = \frac{1}{\sqrt{2}}\ket{0}\bra{1-}, &
B_5 & = \frac{1}{\sqrt{2}}\ket{1}\bra{1-}.
\end{xalignat*}
It is clear that $\Phi_0$ and $\Phi_1$ are both entanglement-breaking
channels, as all of these Kraus operators have rank one
\cite{HorodeckiSR03}.

\section{Sub-optimality of non-adaptive strategies}

We now prove that non-adaptive strategies cannot allow for a perfect
discrimination of the channels $\Phi_0$ and $\Phi_1$ defined in the
previous section, for any finite number $n$ of channel uses.
In more precise terms, we have
\[
\norm{\Phi_0^{\otimes n} - \Phi_1^{\otimes n}}_{\diamond} < 2
\]
for all choices of $n\in\natural$.

We first prove a simpler mathematical fact, which is that there does
not exist a two-qubit density operator $\rho$ for which $\Phi_0(\rho)$
and $\Phi_1(\rho)$ are perfectly distinguishable.
As we will see, the proof is similar when taking the tensor product of
the channel with itself or with an identity channel that acts on an
auxiliary system.
This handles the multiple-copy case with the possible use of an
ancillary space, thus establishing the more general statement above.

Assume toward contradiction that there exists a density operator
$\rho$ such that $\Phi_0(\rho)$ and $\Phi_1(\rho)$ are perfectly
distinguishable.
By a simple convexity argument, we may assume that the same is true
for a pure state $\outer{\psi}$ in place of $\rho$.
In other words, there exists a unit vector $\ket{\psi}$ satisfying
\begin{equation} \label{eq:trace-zero}
\tr\( \, \Phi_1(\outer{\psi}) \, \Phi_0(\outer{\psi}) \, \) = 0 \,.
\end{equation}
Expanding this equation in terms of the Kraus operators of $\Phi_0$
and $\Phi_1$ yields
\[
\sum_{j = 1}^5 \sum_{k = 1}^5
\abs{\bra{\psi} B_j^{\ast} A_k \ket{\psi}}^2 = 0 \,.
\]
Each of the terms in this sum is nonnegative, and must therefore be
zero, i.e., $\bra{\psi} B_j^{\ast} A_k \ket{\psi} = 0$
for all choices of $j,k\in\{1,\ldots,5\}$.
It follows that
\begin{equation} \label{eq:alpha-sum}
\bra{\psi} \sum_{j = 1}^5 \sum_{k = 1}^5 
\alpha_{j,k} B_j^{\ast} A_k \ket{\psi} = 0
\end{equation}
for every choice of complex numbers 
$\{\alpha_{j,k}\,:\,1\leq j,k\leq 5\}$.

We will now obtain a contradiction by choosing the coefficients
$\{\alpha_{j,k}\,:\,1\leq j,k\leq 5\}$ in such a way that
\eqref{eq:alpha-sum} cannot hold.
In particular, by letting
\[
\alpha_{1,1} = \alpha_{2,2} = \sqrt{2},\quad
\alpha_{3,5} = \alpha_{4,3} = 1,\quad\text{and}\quad
\alpha_{4,4} = -2\sqrt{2},
\]
and letting $\alpha_{j,k} = 0$ for all of the remaining values of $j$
and $k$, we find that
\[
\sum_{j = 1}^5 \sum_{k = 1}^5 \alpha_{j,k} B_j^{\ast} A_k = P
\]
for
\[
P = 
\outer{00} +
\outer{01} +
\outer{11} +
\outer{1-}
=
\begin{pmatrix}
1 & 0 & 0 & 0\\
0 & 1 & 0 & 0\\
0 & 0 & 1/2 & -1/2\\
0 & 0 & -1/2 & 3/2
\end{pmatrix}.
\]
The operator $P$ is positive definite and therefore
$\langle \psi | P | \psi \rangle > 0$ for every nonzero vector $\ket{\psi}$,
which is in contradiction with \eqref{eq:alpha-sum}.
Having established a contradiction, we conclude that there cannot
exist a density operator $\rho$ such that $\Phi_0(\rho)$ and
$\Phi_1(\rho)$ are perfectly distinguishable as claimed.

Now let us consider the general setting where an arbitrary finite
number $n$ of (non-adaptive) channel uses, as well as an ancillary
system of arbitrary size, are permitted.
We may follow a similar proof to the one above to show that there
cannot exist a unit vector $\ket{\psi}$ such that
\begin{equation} \label{eq:general-trace-zero}
  \tr\left[ 
    \( \Phi_1^{\otimes n} \otimes \I_{\lin{\W}} \) (\,\outer{\psi}\,)
    \( \Phi_0^{\otimes n} \otimes \I_{\lin{\W}} \) (\,\outer{\psi}\,)
   \right] = 0 \,,
\end{equation}
where $\W$ is the space (of arbitrary finite dimension) that is to be
associated with the ancillary system.
We may express the relevant mappings in this expression in terms of
the Kraus operators of $\Phi_0$ and $\Phi_1$ as follows:
\begin{align*}
  \(\Phi_0^{\otimes n} \otimes \I_{\lin{\W}} \) (X)
  & =
  \sum_{1\leq j_1,\ldots,j_n\leq 5} 
  (A_{j_1}\otimes\cdots\otimes A_{j_n}\otimes \I_{\W}) X
  (A_{j_1}\otimes\cdots\otimes A_{j_n}\otimes \I_{\W})^{\ast},\\
  \(\Phi_1^{\otimes n} \otimes \I_{\lin{\W}} \) (X)
  & =
  \sum_{1\leq j_1,\ldots,j_n\leq 5} 
  (B_{j_1}\otimes\cdots\otimes B_{j_n}\otimes \I_{\W}) X
  (B_{j_1}\otimes\cdots\otimes B_{j_n}\otimes \I_{\W})^{\ast}.
\end{align*}

Now, for the same coefficients $\{\alpha_{j,k}\,:\,1\leq j,k\leq 5\}$
that were defined above, we find that
\[
\sum_{\substack{1\leq j_1,\ldots,j_n\leq 5\\[0.2mm]1\leq k_1,\ldots,k_n\leq 5}}
\alpha_{j_1,k_1} \cdots \alpha_{j_n,k_n} \; 
B_{j_1}^{\ast}A_{k_1} \otimes\cdots\otimes B_{k_n}^{\ast} A_{j_n}
\otimes \I_{\W}
\;=\; P^{\otimes n} \otimes \I_{\W},
\]
which is again positive definite.
Therefore, there cannot exist a nonzero vector $\ket{\psi}$ for which
\[
\bra{\psi}
B_{j_1}^{\ast}A_{k_1} \otimes\cdots\otimes B_{k_n}^{\ast} A_{j_n}
\otimes \I_{\W} \ket{\psi} = 0
\]
for all $j_1,\ldots,j_n$, $k_1,\ldots,k_n$.
Consequently, \eqref{eq:general-trace-zero} does not hold for any
nonzero vector $\ket{\psi}$, which implies that $\Phi_0$ and $\Phi_1$
cannot be perfectly discriminated by means of a non-adaptive strategy.

When the number of evaluations $n$ of the unknown channel is small,
one can efficiently compute the value 
$\norm{\Phi_0^{\otimes n} - \Phi_1^{\otimes n}}_{\diamond}$ because
it is the optimal value of a semidefinite programming problem
\cite{Watrous09}.  
For instance, it holds that
\[
\norm{\Phi_0 - \Phi_1}_{\diamond} = 1 + \frac{1}{\sqrt{2}} \; ,
\]
and therefore the channels can be discriminated with a probability
\[
\frac{1}{2} + \frac{1}{4}\norm{\Phi_0 - \Phi_1}_{\diamond}
\approx 0.9268
\]
of correctness with just a single channel evaluation.
For two non-adaptive queries, we used \texttt{CVX}
\cite{GrantB09,GrantB08}, a package for specifying and solving convex
programs in Matlab, to approximate the value
\[
\frac{1}{2} + \frac{1}{4} \norm{\Phi_0\otimes\Phi_0 -
  \Phi_1\otimes\Phi_1}_{\diamond} \approx 0.9771 \,.
\]
One can also obtain an upper bound on the probability of success using
any feasible solution to the dual problem.  In fact, even obvious
choices give fairly tight upper bounds.
Thus, we establish a small, but finite, advantage of an adaptive strategy
over a non-adaptive one for discriminating these channels.

\section{Remarks on classical channel discrimination}
\label{sec:2classical}

The channels in our example above are entanglement-breaking channels,
yet the optimal adaptive discriminating strategy operates in a
distinctively quantum way: one out of two nonorthogonal key states is
extracted from the first channel evaluation and coherently input to
the second.
A natural question arises, which is whether adaptive strategies also
help when discriminating {\it classical} channels.  
It turns out that adaptive strategies indeed are better in the
classical setting, although in a more limited respect.
This section discusses a few basic facts and examples that illustrate
this claim.

A classical channel can, of course, be succinctly represented by a
stochastic matrix $M$, where the vector $M\ket{k}$ represents the
output distribution when the input is $k$.
Throughout this section, we will let $M_0$ and $M_1$ denote the two
possible channels in a classical channel discrimination problem.

\subsubsection*{Advantages of adaptive classical strategies}

We will present three examples illustrating that adaptive strategies
may give advantages over non-adaptive strategies for classical channel
discrimination, restricting our attention to the special case where
just two channel evaluations are permitted, and where one of two
channels is given with equal probability.
We have the following expression for the optimal success probability
using an adaptive strategy in this setting:
\begin{equation} \label{eq:classical-adaptive}
\frac{1}{2} + \frac{1}{4}\max_{k,f}
\sum_{j} \norm{
M_0(j,k)\,M_0\ket{f(j)} - M_1(j,k)\,M_1\ket{f(j)}}_1.
\end{equation}
In this expression, $j$ and $k$ range over all outputs and inputs,
respectively, of the channels $M_0$ and $M_1$ (i.e., they are row and
column indices).
The function $f$ ranges over all maps from outputs to inputs (or row
indices to column indices).

An alternate expression for the optimal success probability 
\eqref{eq:classical-adaptive} is
\[
\frac{1}{2} + \frac{1}{4}\max_{k}
\sum_{j} q(j,k) \max_{l}
\norm{p_0(j,k)\,M_0\ket{l} - p_1(j,k)\,M_1\ket{l}}_1,
\]
where
\[
q(j,k) = \frac{M_0(j,k) + M_1(j,k)}{2}
\]
and where
\[
p_a(j,k) = \frac{M_a(j,k)}{M_0(j,k)+M_1(j,k)}
\]
is the probability that the unknown channel is $M_a$, conditioned on
$k$ being chosen as the input and $j$ being obtained as the output.
This illustrates that, at least for strategies allowing just two
channel evaluations, that the optimal adaptive strategy for two uses
of a classical channel can be readily found, by first finding the
optimal input for each prior distribution over the chosen channel
(this may be the input in the second use). 
We then compute the success probability given every prior distribution
and one use of the channel. 
Finally, to choose an input to the first use of the channel, we choose
an input which maximizes the probability of getting each prior times the
success probability given that prior. 

\vspace{2mm}\noindent {\bf Example 1.}  
This ``minimal'' example shows that adaptive strategies are better
than nonadaptive ones in some cases.  
The two channels are given by: 
\[
M_0 = \begin{pmatrix} 
  1/3 & 8/9 \\[2mm]
  2/3 & 1/9 
\end{pmatrix},
\quad\quad
M_1 = \begin{pmatrix} 
  0 & 1/3\\[2mm]
  1 & 2/3
\end{pmatrix}.
\]
One can verify that the best two-evaluation non-adaptive strategy is
to input $1$ to both of the channel uses, which leads to a correct
identification with probability $7/9$.
The best adaptive strategy is to take $k = 2$ and 
$f(1) = 2,\;f(2) = 1$ in the formula \eqref{eq:classical-adaptive},
which gives a correct identification with probability $65/81$.
Similar examples are abundant.

\vspace{2mm}\noindent{\bf Example 2.}
Here, the optimal 1-shot input is never used in the optimal
non-adaptive scheme.  
The idea is to start with two optimal $1$-shot inputs $k, k'$ such
that using $k'$ becomes more informative with $2$ parallel uses.  
Then we perturb the $k$-th column slightly so that $k$ becomes the
unique optimal $1$-shot input.  
In this example, the optimal $1$-shot input $k$ still serves as the
first input to the optimal adaptive scheme.

Let the two channels be given by: 
\[
M_0 = \begin{pmatrix} 
0.86 &  0.45 &  1 &  0.5 \\[2mm]
0.14 &  0.1  &  0 &  0.5 \\[2mm]
0    &  0.45 &  0 &  0
\end{pmatrix},
\quad\quad
M_1 = \begin{pmatrix} 
0.15 &  0.1 &  0.5 &  0 \\[2mm]
0.85 &  0.8 &  0.5 &  1 \\[2mm]
0    &  0.1 &  0   &  0
\end{pmatrix}.
\]
The best one-shot input is $k=1$ (probability of success is 0.855)
(whereas $k'= 2$). 
The best parallel input pairs are $(2,3)$ and $(3,2)$ (probability of
success is 0.9).  
Allowing adaptation, and using $k=1$ as the first input, 
$f(1) = 3, f(2) = 4, f(3) = 1$, the probability of success is 0.9275.

\vspace{2mm}\noindent{\bf Example 3.}
In this final example, the optimal $1$-shot input is
{\em not} the first input to the optimal adaptive scheme.  The idea is
to have two optimal $1$-shot inputs in which one is more informative
than the other if given a second use.  Then, we perturb the column
corresponding to the less informative input to be slightly better for
the $1$-shot case.

Let the two channels be given by: 
\[
M_0 = \begin{pmatrix} 
  1 &  0.5 &  0.828 &  0.76\\[2mm]
  0 &  0.5 &  0.092 &  0.04\\[2mm]
  0 &  0   &  0.08  &  0.2
\end{pmatrix},
\quad\quad
M_1 = \begin{pmatrix} 
  0.5 &  0 &  0.092 &  0.04\\[2mm]
  0.5 &  1 &  0.828 &  0.76\\[2mm]
  0   &  0 &  0.08  &  0.2
\end{pmatrix}, 
\]
The best one-shot input is $3$ (probability of success is 0.868) but
the best parallel input pairs to two uses are $(3,4)$ and $(4,3)$
(probability of success is 0.9336).  
The optimal adaptive scheme uses $k=4$ as the first input, and
$f(j) = j$ for $j=1,2,3$, resulting in a probability of success of
0.9536.

\subsubsection*{Perfect classical strategies}

Finally, we give a simple proof of a fact claimed in the introduction
of this paper, which is that if two classical channels are not
perfectly distinguishable with a single evaluation, then they cannot
be perfectly distinguished by any finite number of evaluations, even
using an adaptive strategy.
We will prove the contrapositive of this statement.

\begin{figure}[t]
  \begin{center}
    \unitlength=0.25pt
    \begin{picture}(1600, 200)(-50,0)
      \gasset{Nmr=0,Nw=100}
      \node[Nh=200,fillgray=0.95](in)(0,100){$F_0$}
      \node[Nh=200,Nframe=n](out)(1600,100){}
      \node[Nh=200,fillgray=0.95](K1)(400,100){$F_1$}
      \node[Nh=200,fillgray=0.95](K2)(800,100){$F_2$}
      \node[Nh=200,fillgray=0.95](K3)(1200,100){$F_3$}
      \node[Nh=100,fillgray=0.95](M1)(200,150){$M_a$}
      \node[Nh=100,fillgray=0.95](M2)(600,150){$M_a$}
      \node[Nh=100,fillgray=0.95](M3)(1000,150){$M_a$}
      \node[Nh=100,fillgray=0.95](M4)(1400,150){$M_a$}
      \drawedge[syo=50](in,M1){}
      \drawedge[eyo=50](M1,K1){}
      \drawedge[syo=50](K1,M2){}
      \drawedge[eyo=50](M2,K2){}
      \drawedge[syo=50](K2,M3){}
      \drawedge[eyo=50](M3,K3){}
      \drawedge[syo=50](K3,M4){}
      \drawedge[eyo=50](M4,out){}
      \drawedge[syo=-50,eyo=-50](in,K1){}
      \drawedge[syo=-50,eyo=-50](K1,K2){}
      \drawedge[syo=-50,eyo=-50](K2,K3){}
      \drawedge[syo=-50,eyo=-50](K3,out){}
      \gasset{Nmr=5,Nw=10,Nh=140,ExtNL=y,NLangle=-90,NLdist=20,fillgray=0.95}
      \node(p)(1300,100){$q_a$}
    \end{picture}
  \end{center}
  \caption{The structure of a general discrimination strategy for
    classical channels.
    This example makes four channel evaluations, each illustrated by a
    box labeled $M_a$, but in general any finite number $n$ of
    evaluations may be considered.
    Each arrow represents a register that may be in a random mixture
    over some finite set of classical states, and the boxes labeled
    $F_0$, $F_1$, $F_2$ and $F_3$ represent arbitrary functions (or
    random processes) that must be independent of the value
    $a\in\{0,1\}$ that indicates which of the two channels is given.
  }
  \label{fig:classical-discrimination}
\end{figure}
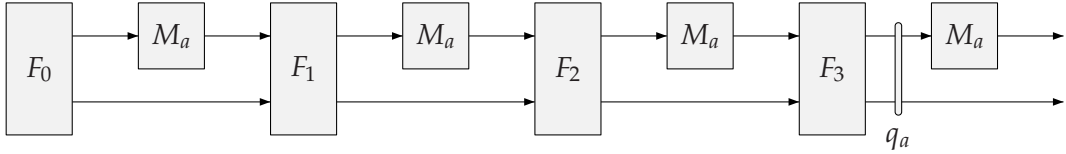

Suppose that two classical channels $M_0$ and $M_1$ are perfectly
discriminated by a discrimination strategy that uses $n$ channel
evaluations.
Without loss of generality we may assume the strategy takes the general
form suggested in Figure~\ref{fig:classical-discrimination}.
The assumption that the strategy perfectly discriminates $M_0$ and
$M_1$ means that the final output distributions for the cases $a=0$
and $a=1$ have disjoint support.
Our goal is to prove that $M_0$ and $M_1$ are perfectly discriminated
with a single evaluation.

The proof of this statement proceeds by induction on $n$.
In case $n=1$ there is nothing to prove, so assume that $n\geq 2$.
Consider the two distributions $q_0$ and $q_1$ that are illustrated in
the figure.
Each distribution $q_a$ represents the state of the discrimination
strategy immediately before the final channel evaluation takes place,
assuming the unknown channel is given by $M_a$.
There are two cases: $q_0$ and $q_1$ have disjoint support, or they do
not.
If $q_0$ and $q_1$ do have disjoint support, then terminating the
discrimination strategy after $n-1$ channel evaluations allows for a
perfect discrimination, so by the induction hypothesis it is possible
to discriminate the channels with a single evaluation.
In the other case, where $q_0$ and $q_1$ do not have disjoint
supports, there must exist a classical state $x$ of the strategy at
the time under consideration for which $q_0(x)$ and $q_1(x)$ are both
positive.
Given that the discrimination strategy is perfect, and therefore has
final distributions with disjoint supports, it must hold that
evaluating $M_0$ and $M_1$ on $x$ results in distributions with
disjoint supports.
Therefore, $M_0$ and $M_1$ can be discriminated with a single channel
evaluation as required.

\section{Conclusion}

In this paper, we presented a pair of quantum channels that can be
discriminated perfectly by a strategy making two adaptive channel
evaluations, but which cannot be perfectly discriminated nonadaptively
with any finite number of channel evaluations.

One natural question that arises is whether our example can be
generalized to show a similar advantage of general adaptive strategies
making $n$ channel evaluations versus strategies that make channel
evaluations with depth at most $n-1$.
Although our example can be generalized in a natural way, we have not
proved that it has the required properties with respect to depth $n-1$
strategies.

Finally, for the example we have presented, we have found that
although strategies making two non-adaptive channel evaluations cannot
be perfect, they can be correct with high probability (about 97.7\%).
What is the largest possible gap between optimal adaptive versus
non-adaptive strategies making two (or any other number of) channel
evaluations?
The only upper-bound we have on this gap is that channels $\Phi_0$ and
$\Phi_1$ that are perfectly discriminated by two (adaptive or
non-adaptive) evaluations must satisfy 
$\norm{\Phi_0 - \Phi_1}_{\diamond} \geq 1$, and can therefore be 
discriminated (with a single evaluation) with probability at least
$3/4$ of correctness.

\subsection*{Acknowledgements}

DL thanks Chris Fuchs for helpful comments.
AWH was funded by the EPSRC grant ``QIP IRC'' and is grateful for the
hospitality of the Perimeter Institute when some of this work was
carried out.
AH received support from the xQIT Keck fellowship.  
DL was funded by CRC, CFI, ORF, CIFAR, NSERC, and QuantumWorks.  
JW was funded by NSERC, CIFAR, and QuantumWorks.


\newcommand{\etalchar}[1]{$^{#1}$}

\end{document}